\documentclass[reprint,amsmath,amssymb,aps,floatfix,groupedaddress,superscriptaddress,nofootinbib,preprintnumbers]{revtex4-1}
\usepackage{float}
\usepackage[pdftex]{hyperref}
\usepackage{graphicx,color}
\usepackage{dcolumn}
\usepackage{bm}
\usepackage{here}
\usepackage{comment}
\usepackage[paperwidth=210mm,paperheight=297mm,centering,hmargin=1.5cm,vmargin=2.0cm]{geometry}
\usepackage{tabularx}

\allowdisplaybreaks[1]

\makeatletter
\newsavebox{\@brx}
\newcommand{\llangle}[1][]{\savebox{\@brx}{\(\m@th{#1\langle}\)}%
  \mathopen{\copy\@brx\mkern2mu\kern-0.9\wd\@brx\usebox{\@brx}}}
\newcommand{\rrangle}[1][]{\savebox{\@brx}{\(\m@th{#1\rangle}\)}%
  \mathclose{\copy\@brx\mkern2mu\kern-0.9\wd\@brx\usebox{\@brx}}}
\makeatother

\begin{document}

\preprint{J-PARC-TH-0118}

\title{A general procedure for detector-response correction
  of higher order cumulants}

\author{Toshihiro Nonaka}
\email{tnonaka@rcf.rhic.bnl.gov}
\affiliation{Key\,Laboratory\,of\,Quark\,\&\,Lepton\,Physics\,(MOE)\,and\,Institute\,of\,Particle\,Physics,\,Central\,China\,Normal\,University,\,Wuhan\,430079,\,China}
\affiliation{Tomonaga\,Center\,for\,the\,History\,of\,the\,Universe,\,University\,of\,Tsukuba,\,Tsukuba,\,Ibaraki\,305,\,Japan}
\author{Masakiyo Kitazawa}
\email{kitazawa@phys.sci.osaka-u.ac.jp}
\affiliation{Department\,of\,Physics,\,Osaka\,University,\,Toyonaka,\,Osaka 560-0043,\,Japan}
\affiliation{J-PARC Branch,\,KEK Theory Center,\,Institute\,of\,Particle\,and\,Nuclear\,Studies,\,KEK,\,203-1,\,Shirakata,\,Tokai,\,Ibaraki,\,319-1106,\,Japan}
\author{ShinIchi Esumi}
\email{esumi.shinichi.gn@u.tsukuba.ac.jp}
\affiliation{Tomonaga\,Center\,for\,the\,History\,of\,the\,Universe,\,University\,of\,Tsukuba,\,Tsukuba,\,Ibaraki\,305,\,Japan}
\date{\today}


\begin{abstract}
We propose a general procedure for the detector-response correction
(including efficiency correction) of higher order cumulants
observed by the event-by-event analysis in heavy-ion collisions.
This method makes use of the moments of the response matrix
characterizing the property of a detector, 
and is applicable to a wide variety of response matrices such as 
those having non-binomial responses and including the effects of 
ghost tracks.
A procedure to carry out the detector-response correction of 
realistic detectors is discussed.
In test analyses, we show that this method can 
successfully reconstruct the cumulants of true distribution
for various response matrices including the one having
multiplicity-dependent efficiency.

\end{abstract}
\maketitle

\newcommand{\ave}[1]{\ensuremath{\langle#1\rangle} }

\section{Introduction}
\label{sec:intro}

Fluctuations are important observables
in relativistic heavy-ion collisions for the search for 
the QCD critical point and the phase transition 
to the deconfined medium~\cite{Asakawa:2015ybt,Luo:2017faz}.
In particular, non-Gaussianity of fluctuations characterized by
the higher order cumulants is believed to be sensitive
to these phenomena~\cite{susceptibility,correlation,Asakawa:2009aj}.
Active measurements of fluctuation observables 
have been performed by the event-by-event
analyses at RHIC~\cite{net_proton,net_charge,Adare:2015aqk,Adamczyk:2017wsl},
the LHC~\cite{Abelev:2012pv,Rustamov:2017lio},
and the NA61~\cite{NA61} and HADES~\cite{HolzmannQM} Collaborations.
Future experimental facilities, J-PARC~\cite{J-PARC-HI},
FAIR~\cite{Rapp:2011zz}, and NICA~\cite{NICA}, will
also contribute this subject.

In relativistic heavy-ion collisions, the cumulants of a particle-number
distribution are obtained from an event-by-event histogram of the
particle number observed experimentally.
However, because of the imperfect capability of detectors
the experimentally-observed event-by-event histogram
is modified from the true distribution, and accordingly
their cumulants are also altered by these artificial effects.
In the experimental analysis, this effect has to be corrected.
In this paper we call this procedure as the detector-response correction.

Compared to standard observables given by expectation values,
the detector-response correction of the cumulants
higher than the first order is more involved,
because the change of a distribution function 
modifies its cumulants in a non-trivial way~\cite{Asakawa:2015ybt}.
So far, the detector-response correction has been discussed by focusing
on the efficiency correction, i.e. the correction for the effects of
the loss of particles at the measurement.
It has been established that the correction can be carried out
if one assumes that the probability that the detector
observes a particle (efficiency) is uncorrelated for individual particles.
In this case, the detector's response is described by the 
binomial distribution~\cite{eff_kitazawa,Asakawa:2015ybt},
that we call the binomial model%
~\cite{Bialas:1985jb,eff_kitazawa,eff_koch,eff_psd_volker,eff_xiaofeng,eff_psd_kitazawa,tsukuba_eff_separate,Nonaka:2017kko}.
Recently, using the method proposed in Ref.~\cite{Nonaka:2017kko},
the analysis of the net-proton number cumulants is realized
up to sixth order~\cite{6th}.

The assumption for the binomial model, however, is 
more or less violated at typical detectors in heavy-ion collisions.
First, the multiplicity dependence of the efficiency of realistic
detectors~\cite{Luo:2015ewa,Adamczyk:2017wsl,HolzmannQM}
suggests the existence of the correlations between efficiencies of
different particles~\cite{binomial_breaking}.
Even worse, typical detectors sometimes measure ghost tracks,
i.e. non-existing particles.
The estimate on systematic uncertainty arising from these effects
is important for reliable experimental analyses of higher order cumulants.

One of the general procedures for the detector-response correction 
is the unfolding method~\cite{DAgostini:1994fjx,Adye:2011gm,Garg:2012nf}.
This method analyzes the true {\it distribution function}
from experimental results and knowledge on the detector.
Strictly speaking, however, the reconstruction of the true
distribution function is an ill-posed problem.
The estimate on the systematic uncertainty of the final results is
a nontrivial task in this method, and a large numerical cost is
required for the iterative analysis to obtain the distribution.
Furthermore, the analysis of the distribution itself seems redundant
because the cumulants are relevant quantities for many purposes.
It is desirable to have a method which enables the analysis of
the cumulants directly without using the distribution.

In the present study, we propose a new method to perform 
the detector-response correction of cumulants directly.
In this method, we relate the moments of true and observed distributions
without using the distribution explicitly.
This method can solve the detector-response correction exactly for 
a wide variety of detector's responses; 
for example, those parametrized by the hypergeometric distribution
and the binomial model with fluctuating probability.
By introducing an approximation with a truncation this method can also
deal with the correction of realistic detectors whose response is
estimated only by Monte Carlo simulations.
We demonstrate by explicit numerical analyses that the correction
by this method can be carried out successfully
for non-binomial response matrices and the response matrix representing 
multiplicity-dependent efficiency.

This paper is organized as follows.
In Sec.~\ref{sec:DRC}, 
we discuss a general procedure for the detector-response correction.
The application of this method to the correction of realistic detectors
is then discussed in Sec.~\ref{sec:practical}.
We perform test analyses of this method in 
Secs.~\ref{sec:test1} and \ref{sec:test2}:
We deal with the exactly solvable cases in Sec.~\ref{sec:test1},
and the multiplicity-dependent efficiency in Sec.~\ref{sec:test2},
respectively.
The last section is devoted to a summary.

\section{Detector-response correction}
\label{sec:DRC}

\subsection{Problem}
\label{sec:problem}

Let us first clarify the problem.
We consider a measurement of the cumulants of
an event-by-event distribution of a particle number $N$,
whose probability distribution is given by $P(N)$.
In experimental measurements, 
typical detectors cannot count the particle number $N$ in each event
accurately due to the miss of particle observation and miscounting
of ghost tracks.
Therefore, the observed particle number $n$ by the detector in an event
is generally different from the true particle number $N$,
and its event-by-event distribution denoted by $\tilde{P}(n)$
is also modified from $P(N)$.
The goal of the detector-response correction is to obtain
the cumulants of $P(N)$ from experimentally-observed distribution 
$\tilde{P}(n)$ and the knowledge on the detector.

To deal with this problem, we assume that 
the probability to observe $n$ particles in an event
with the true particle number $N$ depends only on $N$.
We denote this probability as ${\cal R}(n;N)$;
because ${\cal R}(n;N)$ is a probability, it satisfies 
$\sum_n {\cal R}(n;N)=1$.
We refer to ${\cal R}(n;N)$ as the response matrix.
Using ${\cal R}(n;N)$, $\tilde{P}(n)$ and $P(N)$ are related with
each other as
\begin{align}
  \tilde{P}(n) = \sum_N {\cal R}(n;N) P(N).
  \label{eq:P=RP}
\end{align}
In this study we consider the detector-response correction with
Eq.~(\ref{eq:P=RP}).
Here, the response matrix ${\cal R}(n;N)$ contains
all information on the property of the detector relevant to this problem.
We note that Eq.~(\ref{eq:P=RP}) can describe not only
the efficiency loss, but also the effects of the ghost tracks.

Although the detector-response correction of a single-particle
distribution is considered in Eq.~(\ref{eq:P=RP}),
the correction with multi-particle distributions are usually
necessary in heavy-ion collisions~\cite{Nonaka:2017kko}.
In this study, however, we basically concentrate on 
the single-particle distribution for simplicity.
The generalization of the procedure to multi-particle distributions
is straightforward as discussed in Appendix~\ref{sec:multi}.

\subsection{Notation}
\label{sec:notation}

We denote the $m$th-order moment of
$\tilde{P}(n)$ as
\begin{align}
  \llangle n^m \rrangle = \sum_n n^m \tilde{P}(n),
  \label{eq:<n^m>def}
\end{align}
while that of $P(N)$ is expressed as
\begin{align}
  \langle N^m \rangle = \sum_N N^m P(N). 
  \label{eq:<N^m>def}
\end{align}
The $m$th-order cumulants of these distributions are denoted as
$\llangle n^m \rrangle_{\rm c}$ and $\langle N^m \rangle_{\rm c}$,
respectively.

\subsection{Formal solution}
\label{sec:formal}

Substituting Eq.~(\ref{eq:P=RP}) into Eq.~(\ref{eq:<n^m>def})
one finds that the $m$th-order moment of $\tilde{P}(n)$ is given by 
\begin{align}
  \llangle n^m \rrangle
  &= \sum_N P(N) \sum_n n^m {\cal R}(n;N)
  \nonumber \\
  &= \sum_N P(N) R_m(N) ,
  \label{eq:<n^m>}
\end{align}
where $R_m(N)$ is defined as
\begin{align}
  R_m(N) = \sum_n n^m {\cal R}(n;N).
  \label{eq:R_m}
\end{align}
Because ${\cal R}(n;N)$ with fixed $N$ represents a probability,
$R_m(N)$ is understood as the moments of ${\cal R}(n;N)$.

We then suppose that $R_m(N)$ is expanded as
\begin{align}
  R_m(N) = \sum_{j=0} r_{mj} N^j.
  \label{eq:Taylor}
\end{align}
Substituting Eq.~(\ref{eq:Taylor}) into Eq.~(\ref{eq:<n^m>}),
one obtains
\begin{align}
  \llangle n^m \rrangle = r_{m0} + \sum_{j=1} r_{mj} \langle N^j \rangle.
  \label{eq:<>Taylor}
\end{align}
Equation~(\ref{eq:<>Taylor}) is expressed in the matrix form as
\begin{align}
  \left[\begin{matrix}
    \llangle n \rrangle \\ \llangle n^2 \rrangle \\ \vdots \\
  \end{matrix}\right]
  =
  \left[\begin{matrix}
    r_{10} \\ r_{20} \\ \vdots 
  \end{matrix}\right]
  + \bm{R}
  \left[\begin{matrix}
    \langle N \rangle \\ \langle N^2 \rangle \\ \vdots \\
  \end{matrix}\right],
  \label{eq:nNmat}
\end{align}
with the matrix $\bm{R} = (r_{mj})$.
If $\bm{R}$ is a regular matrix,
by applying $\bm{R}^{-1}$ to Eq.~(\ref{eq:nNmat}) from left 
one obtains
\begin{align}
  \left[\begin{matrix}
      \langle N \rangle \\ \langle N^2 \rangle \\ \vdots \\
  \end{matrix}\right]
  =
  \bm{R}^{-1}
  \left[\begin{matrix}
      \llangle n \rrangle \\ \llangle n^2 \rrangle \\ \vdots \\
  \end{matrix}\right]
  -   \bm{R}^{-1}
  \left[\begin{matrix}
    r_{10} \\ r_{20} \\ \vdots 
  \end{matrix}\right].
  \label{eq:nNmat-inv}
\end{align}

In Eq.~(\ref{eq:nNmat-inv}), the moments of the true distribution
$\langle N^m \rangle$ are expressed by the experimentally-observed
moments $\llangle n^m \rrangle$ together with 
the parameters $r_{mj}$ characterizing the property of the detector.
Because the cumulants of $P(N)$ are obtained from
$\langle N^m \rangle$~\cite{Asakawa:2015ybt},
the detector-response correction of the cumulants is carried out 
with Eq.~(\ref{eq:nNmat-inv}).

In this argument, one can use alternative expansions of $R_m(N)$
instead of Eq.~(\ref{eq:Taylor}), which lead to the same final
result but might lead to an enhancement of numerical stability.
This strategy is discussed in Appendix~\ref{sec:expansion}.

\subsection{Exactly solvable models}
\label{sec:exact}

Although the above argument formally solves the 
problem of the detector-response correction, 
Eq.~(\ref{eq:nNmat-inv}) is not quite useful when
the matrix $\bm{R}$ is not closed at some order.
First of all, the inverse matrix of $\bm{R}$ is not
determined in general in this case.
Second, even if the inverse matrix were obtained, 
Eq.~(\ref{eq:nNmat-inv}) requires
$\llangle n^m \rrangle$ up to infinitely higher orders.
In realistic situations, however, the moments $\llangle n^m \rrangle$
accessible with a reasonable statistics are limited
typically to $m\lesssim 6$. 

When the expansions Eq.~(\ref{eq:Taylor}) are terminated at finite orders,
Eq.~(\ref{eq:nNmat-inv}) gives a closed form and provides
exact formulas for the detector-response correction.
Particularly important examples of ${\cal R}(n;N)$
satisfying this condition are the cases that 
$R_m(N)$ is given by a $m$th-order polynomial, i.e.
\begin{align}
  R_m(N) = \sum_{j=0}^{m} r_{mj} N^j.
  \label{eq:Taylor-m}
\end{align}
In this case, the matrix $\bm{R}$ in Eq.~(\ref{eq:nNmat})
has a lower-triangular form
\begin{align}
  \bm{R} = 
  \left[\begin{matrix}
      r_{11} & 0 & 0 & \cdots \\
      r_{21} & r_{22} & 0 & \cdots\\
      r_{31} & r_{32} & r_{33} & \ddots \\
      \vdots &  \vdots & & \ddots 
  \end{matrix}\right].
  \label{eq:Rr}
\end{align}
The inverse of a lower-triangular matrix is obtained order-by-order,
and $\bm{R}^{-1}$ is also lower triangular.
Substituting the lower-triangular form of $\bm{R}^{-1}$ into
Eq.~(\ref{eq:nNmat-inv}),
one finds that $\langle N^m \rangle$ depends only on 
$\llangle n^l \rrangle$ for $l\le m$.

The binomial model with
${\cal R}_{\rm bin}(n;N)=B(n;p,N)$~\cite{Asakawa:2015ybt},
with the binomial distribution 
\begin{align}
  B(n;p,N) = p^n (1-p)^{N-n} \frac{N!}{n! (N-n)! }
  \label{eq:binomial}
\end{align}
corresponds to this case.
In fact, all the cumulants of the binomial distribution is
proportional to $N$,
\begin{align}
  \langle n^m \rangle_{\rm c,binomial} = \xi_m(p) N ,
  \label{eq:<n^m>B}
\end{align}
where the coefficients $\xi_m(p)$ depend only on $p$~\cite{Nonaka:2017kko}.
Converting Eq.~(\ref{eq:<n^m>B}) into moments,
one immediately finds that $R_m(N)$ in this case is given by
a $m$th-order polynomial as in Eq.~(\ref{eq:Taylor-m}).
In this case, Eq.~(\ref{eq:nNmat-inv}) reproduces the formulas of
the efficiency correction in the binomial model~\cite{Asakawa:2015ybt}.

Other examples satisfying Eq.~(\ref{eq:Taylor-m}) are the binomial model
but the probability $p$ is fluctuating event by event~\cite{He:2018mri},
\begin{align}
  {\cal R}_{\rm G}(n;N)= \int_0^1 dp G(p) B(n;p,N),
  \label{eq:R=GB}
\end{align}
where $G(p)$ is a probability distribution satisfying 
$\int_0^1 dp G(p)=1$.
The moments $R_m(N)$ of Eq.~(\ref{eq:R=GB}) satisfy
Eq.~(\ref{eq:Taylor-m}) for arbitrary forms of $G(p)$,
as discussed in Appendix~\ref{sec:binomial-fluc}.
The detector-response correction for ${\cal R}_{\rm G}(n;N)$
thus is handled with Eq.~(\ref{eq:nNmat-inv}) exactly.
The beta-binomial distribution, which is obtained by 
$G(p)={\cal B}(p;a,b)$ with the beta distribution 
${\cal B}(p;a,b)$, belongs to this case
(see Appendix~\ref{sec:HG-beta}).

Another interesting response matrix is the one parametrized
by the hypergeometric distribution as
\begin{align}
  {\cal R}_{\rm HG}(n;N) = H(n;N,X,Y),
  \label{eq:R=HGD}
\end{align}
where the hypergeometric distribution $H(n;N,X,Y)$ is defined
in Appendix~\ref{sec:HG-beta}.
As shown in Appendix~\ref{sec:HG-beta},
the moments of Eq.~(\ref{eq:R=HGD}) is given in the form 
in Eq.~(\ref{eq:Taylor-m}).
Therefore, the detector-response correction for Eq.~(\ref{eq:R=HGD})
is also carried out exactly with Eq.~(\ref{eq:nNmat-inv}).

\subsection{Truncation}
\label{sec:truncation}

When the expansion of $R_m(N)$ is not closed, one must introduce
an approximation to deal with the detector-response correction.
A simple approximation is a truncation of the expansion
Eq.~(\ref{eq:Taylor}) at $L$th order,
\begin{align}
  R_m(N) = \sum_{j=0}^L r_{mj} N^j.
  \label{eq:TaylorL}
\end{align}
Using Eq.~(\ref{eq:TaylorL}), one obtains a closed formula
up to the $L$th order
\begin{align}
  \left[\begin{matrix}
    \llangle n \rrangle \\ \llangle n^2 \rrangle \\ \vdots \\ \llangle n^L \rrangle 
  \end{matrix}\right]
  =
  \left[\begin{matrix}
    r_{10} \\ r_{20} \\ \vdots \\ r_{L0}
  \end{matrix}\right]
  + \bm{R}
  \left[\begin{matrix}
    \langle N \rangle \\ \langle N^2 \rangle \\ \vdots \\ \langle N^L \rangle
  \end{matrix}\right],
  \label{eq:nNmatL}
\end{align}
where $\bm{R}$ is a $L\times L$ matrix.
When $\bm{R}$ is a regular matrix, by applying $\bm{R}^{-1}$ from left
Eq.~(\ref{eq:nNmatL}) enables one to 
carry out the detector-response correction up to the $L$th order
using the experimental data on $\llangle n^m \rrangle$ for $m\le L$.

Of course, this analysis can be justified only when
the truncated formula Eq.~(\ref{eq:TaylorL}) well reproduces
the functional form of $R_m(N)$.
When ${\cal R}(n;N)$ is given by an analytic form,
the effect of the truncation would be estimated analytically.
When one considers the response matrices of realistic detectors, 
they are usually estimated by Monte Carlo simulations such as
GEANT~\cite{GEANT}, which provide 
the moments $R_m(N)$ with statistical errors.
In this case, one may perform fits to $R_m(N)$ with
Eq.~(\ref{eq:TaylorL}).
The use of Eq.~(\ref{eq:nNmatL}) would be justified as long as
these fits reproduce $R_m(N)$ within statistics.
The detector-response correction of realistic detectors 
will be discussed in Sec.~\ref{sec:practical} in more detail.

\section{Practical analysis}
\label{sec:practical}

In this section, we discuss the detector-response correction
of realistic detectors whose
response matrix ${\cal R}(n;N)$ are not given by an analytic form.
In the following, we consider the use of the approximation with
the truncation discussed in Sec.~\ref{sec:truncation}.

The form of ${\cal R}(n;N)$  of realistic detectors is usually
estimated by Monte Carlo simulations such as GEANT~\cite{GEANT}.
The simulations provide the moments $R_m(N)$ with statistical errors.
In this case, the coefficients $r_{mj}$ in Eq.~(\ref{eq:TaylorL})
are determined by the fits to $R_m(N)$ obtained by the simulation.
Using $r_{mj}$ thus obtained,
the correction can be carried out with Eq.~(\ref{eq:nNmatL}).

Because we do not know the true distribution of $N$
in realistic situations, in the Monte Carlo simulations
one may assume a presumed ``true'' distribution $P_{\rm MC}(N)$.
A problem here is that 
the quality and result of the fits to $R_m(N)$ depend on the form
of $P_{\rm MC}(N)$ and the number of the Monte Carlo events,
$N_{\rm event}$.
The validity of the fits would be checked by setting 
$N_{\rm event}$ to the same value as the statistics of the experimental data.
When the value of chi-square, $\chi^2/{\rm ndf}$, of these fits are
close to unity with this statistics,
there are no reasons to reject the use of Eq.~(\ref{eq:nNmatL}).
Next, the fitting results of $r_{mj}$ can 
also depend on the form of $P_{\rm MC}(N)$.
This suggests that one must check the sensitivity of the fit results
on the form of $P_{\rm MC}(N)$, or perform an iterative procedure as follows:
\begin{enumerate}
\item 
  Generate ${\cal R}(n;N)$ by a Monte-Carlo simulation with
  a presumed distribution $P_{\rm MC}(N)$.
\item
  Perform fits to $R_m(N)$ with Eq.~(\ref{eq:TaylorL}).
  One then obtains $r_{mj}$ for $m,j \le L$.
  Together with the experimental results on $\llangle n^m \rrangle$,
  one obtains the corrected moments $\langle N^m \rangle$.
\item
  If $\langle N^m \rangle$ thus obtained have large deviations from 
  the moments of $P_{\rm MC}(N)$, replace $P_{\rm MC}(N)$ with the one 
  consistent with $\langle N^m \rangle$ obtained in the above step,
  and take the analysis from the top again.
\item
  Repeat this iteration until $P_{\rm MC}(N)$ is consistent 
  with $\langle N^m \rangle$ obtained by the correction.
\end{enumerate}
It, however, is expected that the result of the fits are insensitive
to $P_{\rm MC}(N)$, especially on the cumulants higher than the second order.
The use of the Gaussian distribution with 
the mean and variance obtained by the correction 
for $P_{\rm MC}(N)$ would be sufficient for this analysis.
It is also expected that a few iterations are enough for convergence.

Finally, we comment on the error analysis.
First, in the detector-response correction with Eq.~(\ref{eq:nNmatL}),
it is important to reflect the correlation between the errors of $r_{mj}$
to the final result appropriately.
An automatic way to include the correlation is
the use of the bootstrap or jackknife analysis
with the successive generation of Monte Carlo events.
Second, in the present method 
it is possible to reduce the errors of $r_{mj}$ by increasing
$N_{\rm event}$ independently of the statistics of $\llangle n^m \rrangle$.
In fact, in the next section we will see that the suppression of
the error of $r_{mj}$ is effective in reducing the error of the final result.
With increasing $N_{\rm event}$, however, the $\chi^2/{\rm ndf}$ of the
fits to $R_m(N)$ with Eq.~(\ref{eq:TaylorL}) will eventually
become unacceptably large.
In this case, the analysis with the truncation loses its validity.
In this sense, this analysis has an upper limit of the resolution.
Third, the effect of the truncation can be estimated 
by comparing the corrected results at the $L$ and $(L+1)$th orders.
Such analyses would require large statistics, but are desirable
for a proper estimate on the systematic uncertainty of the analysis.

\section{Test analysis 1: Exact models}
\label{sec:test1}

In this and next sections, we perform test analyses for the
detector-response correction discussed in Sec.~\ref{sec:DRC}
with toy models for ${\cal R}(n;N)$, and show that 
the corrections are carried out successfully in these cases.

In this section, we first perform test analyses for the response
matrices which can be solved exactly discussed in Sec.~\ref{sec:exact}.
We consider two non-binomial models for ${\cal R}(n;N)$
parametrized by the hypergeometric and beta-binomial distributions as
\begin{align}
  {\cal R}_{\rm HG}(n;N) &= H(n;N,X,Y),
  \label{eq:R_HGD}\\
  {\cal R}_\beta(n;N) &= \beta(n;N,X,Y-X),
  \label{eq:R_beta}
\end{align}
where the hypergeometric and beta-binomial distributions,
$H(n;N,X,Y)$ and $\beta(n;N,a,b)$, are defined in Appendix~\ref{sec:HG-beta}.
The response matrices parametrized by these distributions are
studied in Ref.~\cite{binomial_breaking}
as examples that the binomial model fails in obtaining the true cumulants,
and are good starting points for the check of the new method.
Equations~(\ref{eq:R_HGD}) and (\ref{eq:R_beta})
approach the binomial model ${\cal R}_{\rm bin}(n;N)=B(n;p,N)$ in the
$Y\to\infty$ limit with fixed $p=X/Y$, while
the distribution of $n$ in ${\cal R}_{\rm HG}(n;N)$ (${\cal R}_\beta(n;N)$)
is narrower (wider) than the binomial distribution with finite $Y$.
As discussed in Appendix~\ref{sec:HG-beta},
the values of $r_{mj}$ in Eq.~(\ref{eq:Taylor})
are obtained analytically for ${\cal R}_{\rm HG}(n;N)$ and
${\cal R}_\beta(n;N)$.

\begin{figure} 
  \begin{center}
    \includegraphics[width=0.49\textwidth]{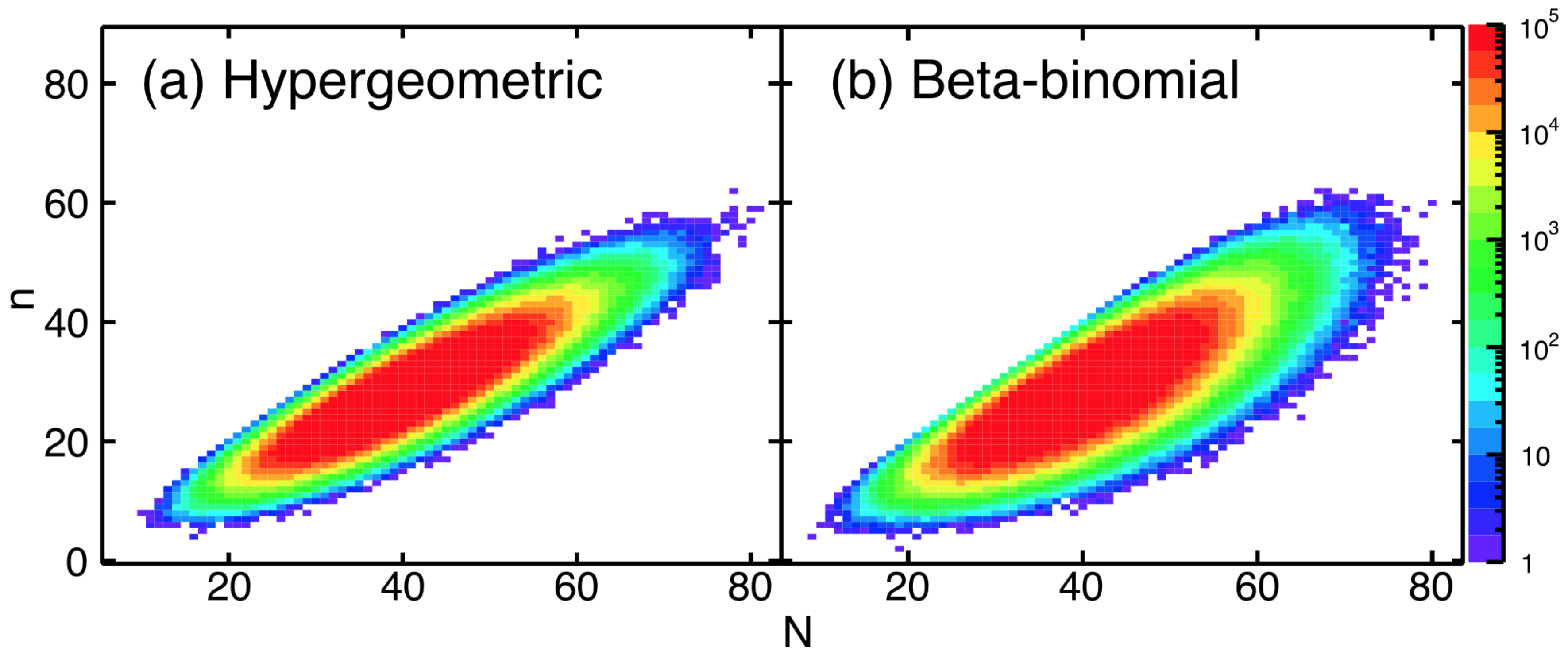}
  \end{center}
  \caption{ Correlation between $n$ and $N$ on the sample events,
    i.e. the magnitude of ${\cal R}(n;N)P(N)$, 
    for the response matrices ${\cal R}_{\rm HG}(n;N)$ (hypergeometric)
    and ${\cal R}_\beta(n;N)$ (beta-binomial) with 
    $p=X/Y=0.7$ and $Y=140$. 
}
  \label{fig:rm}
\end{figure}

The procedure of the test analysis is as follows. 
We first generate sample events of $N$ by assuming 
the Poisson distribution for $P(N)$ with $\langle N \rangle=40$.
We then specify the value of $n$ for each sample event 
randomly according to the probability ${\cal R}_{\rm HG}(n;N)$ or
${\cal R}_\beta(n;N)$.
This allows one to obtain the moments $\llangle n^m \rrangle$.
These moments are used for the correction in Eq.~(\ref{eq:nNmat-inv}).
To proceed the correction, 
we take the following two different analyses.
First, because the values of $r_{mj}$ are analytically known for
${\cal R}_{\rm HG}(n;N)$ and ${\cal R}_\beta(n;N)$,
we perform the correction with these values.
Besides this analysis, as a second option, 
we analyze $\langle N^m \rangle$ with the values of $r_{mj}$ determined by 
the fits to $R_m(N)$ obtained on the sample events with statistical errors.
The second analysis supposes the correction of realistic detectors,
of which the response matrix is obtained only stochastically.

In Fig.~\ref{fig:rm}, we show the correlation between 
$n$ and $N$ on the $10^8$ sample events by plotting the two-dimensional
histogram as a function of $n$ and $N$ for 
the hypergeometric (${\cal R}_{\rm HG}(n;N)$) and beta-binomial
(${\cal R}_\beta(n;N)$) distributions with $p=0.7$ and $Y=140$.
(This plot thus represents the magnitude of ${\cal R}(n;N)P(N)$, 
and is usually called the ``response matrix'' in literature
for simplicity.)
One finds from the figure that the distributions are clearly different
between the two response matrices; the width of $n$ with fixed $N$ is
narrower for ${\cal R}_{\rm HG}(n;N)$ than ${\cal R}_\beta(n;N)$.

\begin{figure} 
  \begin{center}
    \includegraphics[width=0.49\textwidth]{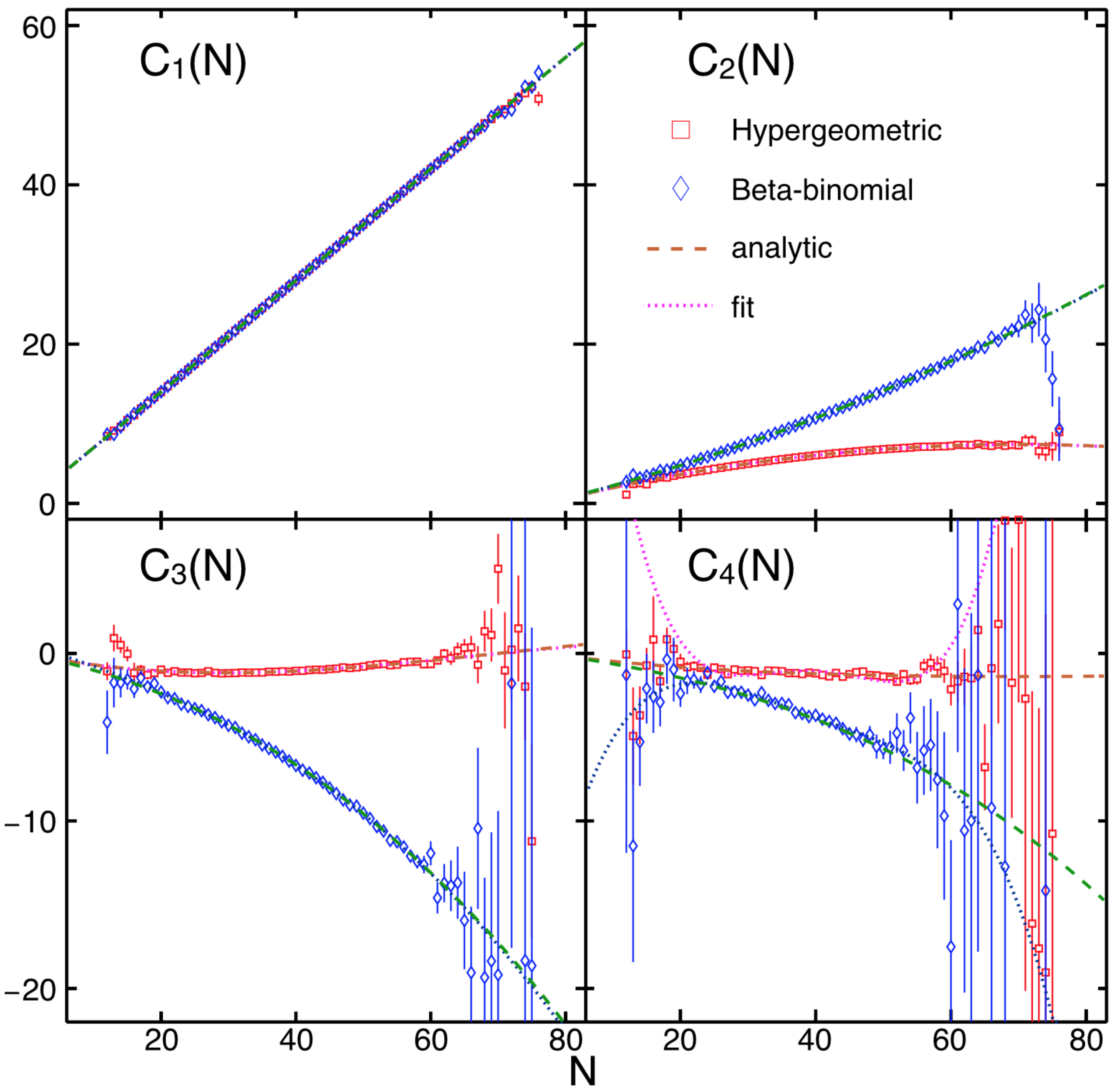}
  \end{center}
  \caption{
    Cumulants of the response matrix $C_m(N)$
    for ${\cal R}_{\rm HG}(n;N)$ and ${\cal R}_\beta(n;N)$
    obtained on $10^8$ sample events with $p=0.7$ and $Y=140$.
    The dashed lines show the analytic values, while the dotted lines
    represent the fitting results with $m$th-order polynomial.
  }
  \label{fig:rm_cum}
\end{figure}

In Fig.~\ref{fig:rm_cum}, we show the cumulants of the
response matrix $C_m(N)$ defined by
\begin{align}
  C_1(N) = R_1(N), \quad C_2(N) = R_2(N) - (R_1(N))^2,
  \label{eq:C}
\end{align}
and so forth,
for ${\cal R}_{\rm HG}(n;N)$ and ${\cal R}_\beta(n;N)$
obtained on $10^8$ sample events with $p=0.7$ and $Y=140$
for $m\le4$.
The dashed lines show the analytic values,
while the dotted lines
are the fitting results with the $m$th-order polynomial.
From these fits one obtains the values of $r_{mj}$.

\begin{figure} 
  \begin{center}
    \includegraphics[width=0.49\textwidth]{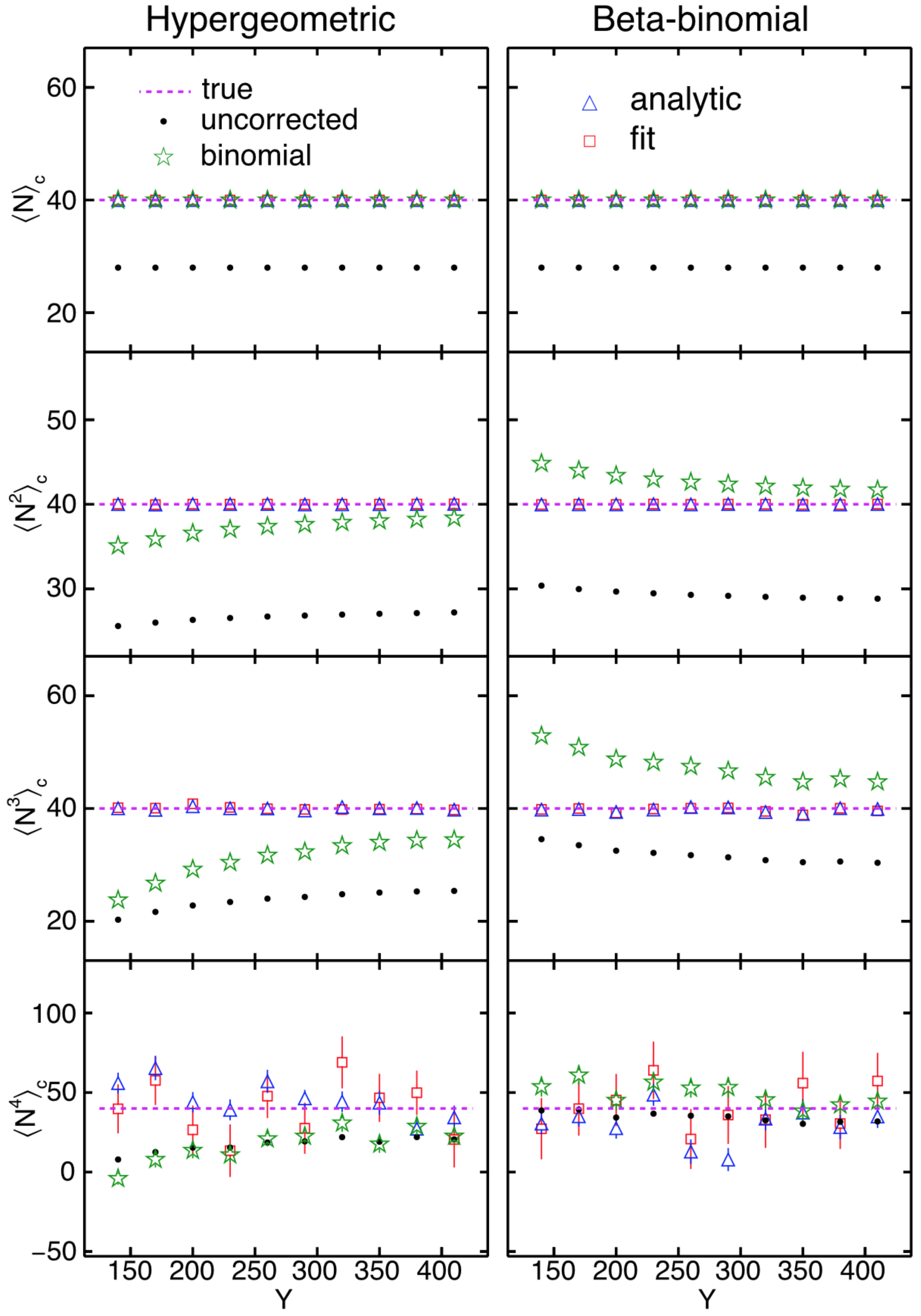}
  \end{center}
  \caption{
    Cumulants obtained by the detector-response correction,
    $\langle N^m \rangle$,
    up to the fourth order with $p=0.7$ as functions of $Y$ for
    ${\cal R}_{\rm HG}(n;N)$ (left) and ${\cal R}_\beta(n;N)$ (right).
    The results obtained with the analytic (fitted) values of $r_{mj}$ are
    shown by triangles (squares).
    The corrected values agree with the true cumulants
    $\langle N^m \rangle_{\rm c}=40$ shown by the dashed line within statistics.
    The uncorrected cumulants $\llangle n^m \rrangle_{\rm c}$ and
    the corrected results in the binomial model are also shown by
    circles and stars, respectively.
  }
  \label{fig:toy_nonbinom}
\end{figure}

In Fig.~\ref{fig:toy_nonbinom}, we show the corrected values of the
cumulants $\langle N^m \rangle_{\rm c}$ for $m\le4$
with $p=0.7$ and various values of $Y$.
The left (right) panel shows the results for ${\cal R}_{\rm HG}(n;N)$
(${\cal R}_\beta(n;N)$).
The triangles represent the results obtained 
with the analytic values of $r_{mj}$, while the results obtained
with $r_{mj}$ determined by the fits to $R_m(N)$ are shown by squares.
$10^7$ sample events are used to obtain $\llangle n^m \rrangle$ in
both analyses, while $r_{mj}$ in the latter analysis are obtained
with $10^8$ sample events.
Errors are estimated by repeating the same simulation $100$ times.
One finds from the figure that the corrected cumulants
$\langle N^m \rangle_{\rm c}$ are consistent with the true value,
$\langle N^m \rangle_{\rm c}=40$ shown by the dashed line,
within statistics for all values of $Y$ in both analyses.
In Fig.~\ref{fig:toy_nonbinom}, the uncorrected cumulants,
$\llangle n^m \rrangle_{\rm c}$, are shown by filled circles.
We also show the results of the efficiency correction with 
the binomial model with $p=0.7$ by the stars.
The results in the binomial model fail in reproducing
the true cumulants~\cite{binomial_breaking}, in contrast to
the new method.

From Fig.~\ref{fig:toy_nonbinom} one also finds 
that the statistical error is large when $r_{mj}$ are determined
by the fits, although the statistics to determine $r_{mj}$ 
is one order larger than that for $\llangle n^m \rrangle$.
This suggests that the suppression of the uncertainty of 
$r_{mj}$ is crucial in reducing the error of the final results.

Finally, we note that the fitting results of $C_m(N)$
in Fig.~\ref{fig:rm_cum} have significant deviations from
the analytic values for $N\gtrsim60$.
Nevertheless, the final results obtained with 
these fits reproduce the true values within statistics.
This result shows that the detector-response correction 
is carried out appropriately even if the fits do not
reproduce $R_m(N)$ in the range of $N$ at which $P(N)$ is small.

\section{Test analysis 2: Multiplicity-dependent efficiency}
\label{sec:test2}

Next, we perform a test analysis of the detector-response correction
for the response matrix which cannot be solved exactly.
As such an example, we consider the response of a detector 
having a multiplicity-dependent efficiency.
We consider the binomial distribution but
the efficiency is dependent on $N$, i.e.
\begin{align}
  {\cal R}_{\rm MD}(n;N) = B(n;p(N),N).
  \label{eq:R_MD}
\end{align}
In typical detectors, the efficiency decreases with increasing
multiplicity ($N$)~\cite{Adamczyk:2017wsl,HolzmannQM}.
To model this behavior we assume 
\begin{align}
  p(N) = 0.7 - \varepsilon(N-\langle N \rangle),
  \label{eq:p(N)}
\end{align}
with $\varepsilon>0$\footnote{
  The distribution of $n$ of realistic detectors with fixed $N$
  would be wider or narrower than the binomial distribution.
  To model such a behavior with the multiplicity-dependent efficiency
  Eq.~(\ref{eq:p(N)}), one may, for example, employ
  the response matrix ${\cal R}(n;N)=\beta(n;N,p(N)Y,Y-p(N)Y)$ or
  ${\cal R}(n;N)=H(n;N,p(N)Y,Y)$.
}

One can analytically show that the $m$th-order moment $R_m(N)$ of
${\cal R}_{\rm MD}(n;N)$ with Eq.~(\ref{eq:p(N)}) 
is given by the $2m$th-order polynomial.
Therefore, the detector-response correction in this case 
cannot be solved exactly by the procedure in Sec.~\ref{sec:exact}.
In the following, we use the truncated formulas 
discussed in Sec.~\ref{sec:truncation} with $L=4$.

\begin{figure} 
  \begin{center}
    \includegraphics[width=0.49\textwidth]{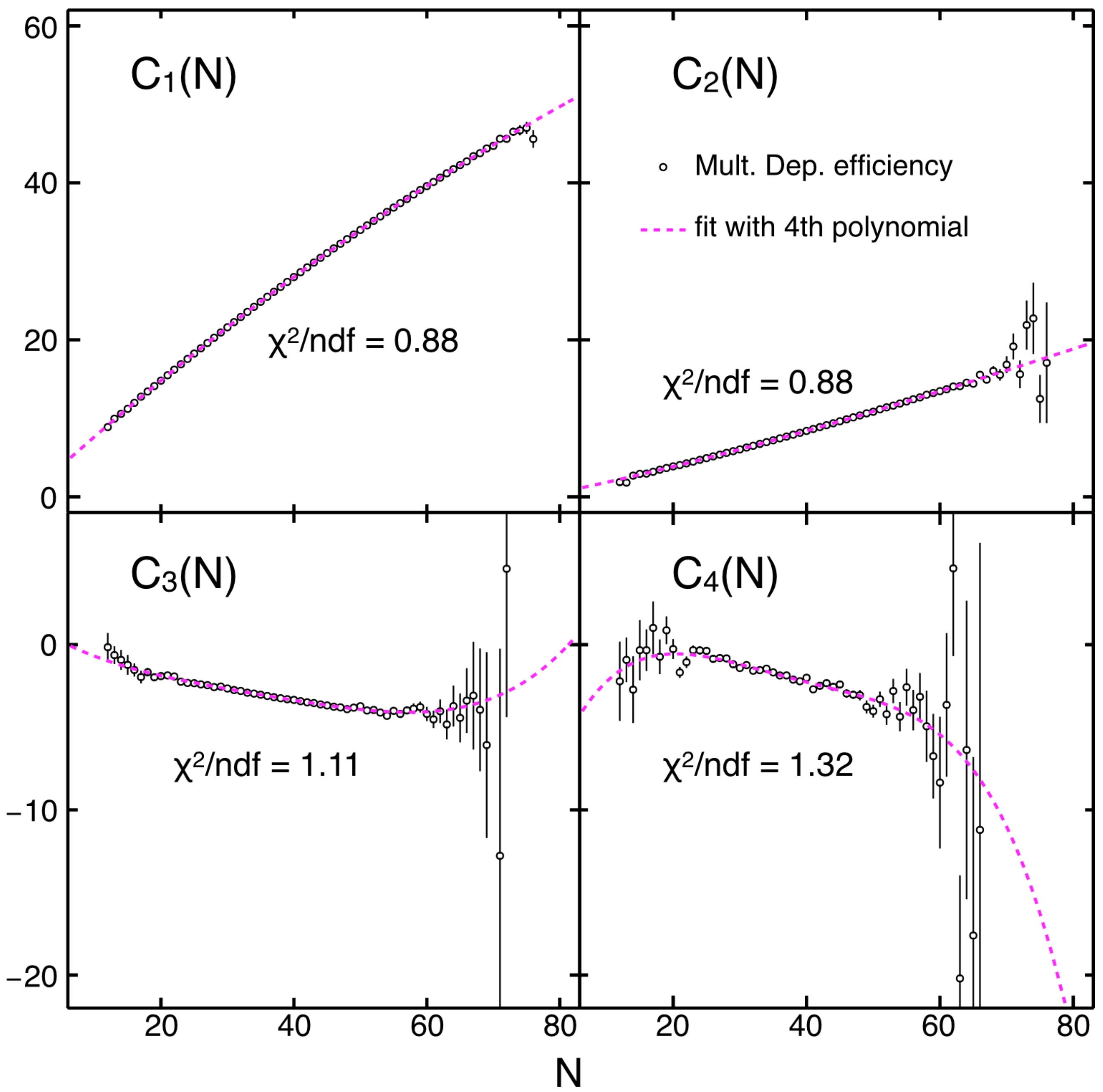}
  \end{center}
  \caption{
    Cumulants $C_m(N)$ of the response matrix $R_{\rm MD}(n;N)$
    with $p=0.7$ and $\varepsilon=0.002$ up to the fourth order
    obtained on $10^8$ sample events.
    The dashed lines show the results of the fits by the fourth order
    polynomial.
  }
  \label{fig:rm_slope}
\end{figure}

\begin{figure} 
  \begin{center}
    \includegraphics[width=0.49\textwidth]{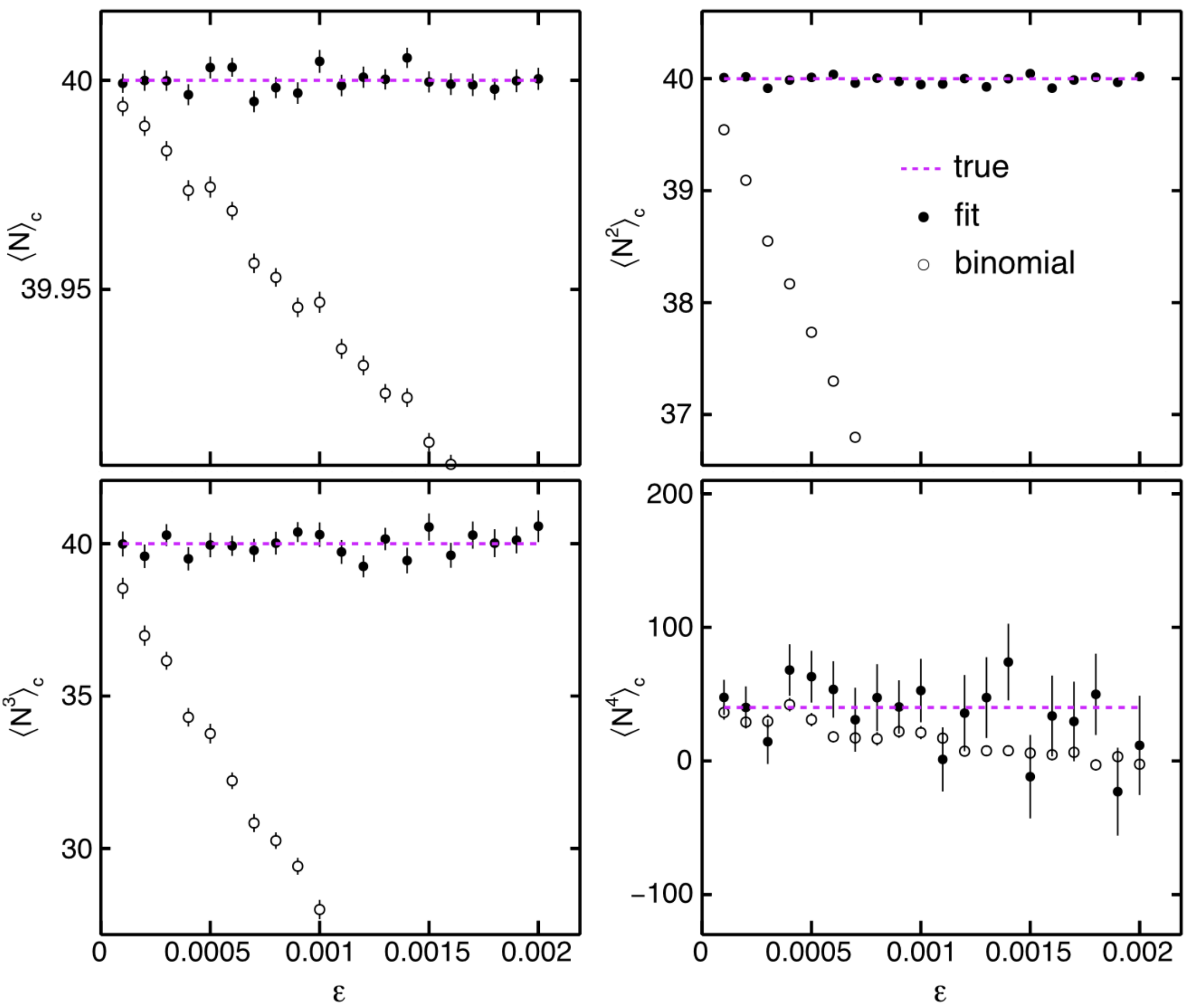}
  \end{center}
  \caption{ 
    Filled circles show the corrected cumulants
    $\langle N^m\rangle_{\rm c}$ obtained by the truncation analysis
    for $R_{\rm MD}(n;N)$ as functions of $\varepsilon$.
    The corrected cumulants reproduce the true value
    $\langle N^m\rangle_{\rm c}=40$ for all $\varepsilon$ within statistics.
    The open circles show the corrected results in the binomial model.
  }
  \label{fig:slope}
\end{figure}

For the test analysis, we generate the sample events of $N$
assuming the Poisson distribution with $\langle N \rangle=40$
for $P(N)$.
The value of $n$ in each sample event is then specified
according to ${\cal R}_{\rm MD}(n;N)$.
$10^8$ sample events are generated in this way.
Figure~\ref{fig:rm_slope} shows the cumulants of the response matrix,
$C_m(N)$, for $m\le4$ obtained from these sample events with
$p=0.7$ and $\varepsilon=0.002$.
We perform the fits to $R_m(N)$ with the fourth-order polynomial
Eq.~(\ref{eq:TaylorL}).
The results of the fits in terms of $C_m(N)$ are shown
by the dashed lines in Fig.~\ref{fig:rm_slope}\footnote{
  We also tested the fits to $C_m(N)$, instead of $R_m(N)$, by 
  Eq.~(\ref{eq:TaylorL}).
  We have checked that the results converted to $R_m(N)$ agrees
  with those obtained by the fit to $R_m(N)$ directly to a good accuracy.
  }.
As shown in the figure, 
we have $\chi^2/{\rm ndf}\simeq1$ for these fits,
which suggests that the fits work well.

The results of the corrected cumulants, $\langle N^{m} \rangle_{\rm c}$,
are shown in Fig.~\ref{fig:slope} by filled circles 
as a function of $\varepsilon$ for $m\le4$.
Errors are estimated by the same procedure as in the previous section.
The figure shows that the corrected results reproduce 
$\langle N^m \rangle_{\rm c}=40$ shown by the dashed line in each panel
within statistics for all values of $\varepsilon$.
In Fig.~\ref{fig:slope}, we also show the cumulants obtained
by the efficiency correction in the binomial model by the open circles.
These results fail in reproducing the true cumulants even at $m=1$.
Although the results in the binomial model are consistent with
the true value for $m=4$, this agreement would be accidental.
We note that the typical value of $\varepsilon$ for protons
at the STAR detector is 
$\varepsilon\simeq0.0002$ at $\sqrt{s_{\rm NN}}=7.7~$GeV,
and $\varepsilon\simeq0.0003$
at $\sqrt{s_{\rm NN}}=200~$GeV in Au+Au collisions
~\footnote{
We note that these numbers are the slope with respect 
to the multiplicity used for the centrality definition, 
which are taken into account in the experimental analysis 
at STAR by using the Centraltiy Bin Width Correction~\cite{Luo:2017faz}. 
There is no study that discusses the slope with respect 
to the net-particle itself, yet.
} 
~\cite{nonaka_WPCF}. 
The maximal value of $\varepsilon$ in Fig.~\ref{fig:slope} is
about one order larger than this value.


As discussed in Sec.~\ref{sec:practical},
in realistic experimental analysis we do not know the
true distribution $P(N)$.
This means that the distribution $P_{\rm MC}(N)$ used in the Monte Carlo
simulation to determine ${\cal R}(n;N)$ is in general different from
$P(N)$.
In order to test the detector-response correction in such a case, 
we performed one more test analysis as follows:
We use the Poisson distribution with $\langle N \rangle=40$
for $P(N)$ to determine $\llangle n^m \rrangle$ as in the above analysis,
but the values of $r_{mj}$ are determined by the fits to $R_m(N)$
generated by the Gauss distribution for $P_{\rm MC}(N)$ with 
$\langle N \rangle=\langle N^2 \rangle_{\rm c}=40$.
We checked that the final results in this analysis agrees with
those shown in Fig.~\ref{fig:slope} within statistics even in this case.
This result suggests that the detector-response correction with
the strategy in Sec.~\ref{sec:practical} is applicable to
realistic analyses.

\section{Summary}
\label{sec:summary}

In this paper, we proposed a new procedure to carry out the detector-response
correction (including efficiency correction) of the higher order cumulants
in the event-by-event analysis.
This method provides exact formulas for various non-binomial
response matrices, including Eqs.~(\ref{eq:R_HGD}) and
(\ref{eq:R_beta}) parametrized 
by the hypergeometric and beta-binomial distributions.
Introducing an approximation with the truncation, this method
can deal with the correction of realistic detectors.
The correction in this method is demonstrated explicitly for 
three non-binomial response matrices.
We showed that the true cumulants are obtained within
statistics not only for the exactly-solvable cases but also a case
that the approximation with the truncation is necessary.
Although we concentrated on the correction for the single-variable
distribution throughout this manuscript, the extension to deal with 
the multi-variable distribution is also possible as discussed
in Appendix~\ref{sec:multi}.'' ``

We thank M.~Gazdzicki, R.~Holzmann, A.~Rustamov, M.~Szala, and N.~Xu
for constructive discussions.
M.~K. thanks K.~Redlich and B.~Friman for inviting him to
the workshop ``Constraining the QCD Phase Boundary with Data
from Heavy Ion Collisions'' (Feb. 12-14, 2018, Darmstadt, Germany)
and fruitful discussions during the workshop.

\appendix

\section{Detector-response correction for multi-variable distribution}
\label{sec:multi}

In this appendix, we discuss the extension of the
detector-response correction to multi-variable distribution functions.
This extension is necessary for the analysis of the net-particle
number cumulants, because the net number is given by the difference
of particle and anti-particle numbers.
Also, the effects of the momentum and azimuthal angle dependence of
the detector's response can be described by the multi-variable
distribution~\cite{Nonaka:2017kko}.

For a simple illustration, we consider
a distribution function of two particle species, $P( N_1, N_2 )$, 
where $N_1$ and $N_2$ are the numbers of the particles in each event.
We then denote the numbers of observed particles in each event
as $n_1$ and $n_2$, respectively, and its event-by-event
distribution as $\tilde{P}( n_1, n_2 )$.

Similar to Eq.~(\ref{eq:P=RP}), we assume that these distribution
functions are related with each other as
\begin{align}
  \tilde{P}(n_1,n_2) = \sum_{N_1,N_2} {\cal R}(n_1,n_2;N_1,N_2) P(N_1,N_2),
  \label{eq:P=RPmult}
\end{align}
where the response matrix ${\cal R}(n_1,n_2;N_1,N_2)$ satisfies
the normalization condition $\sum_{n_1,n_2} {\cal R}(n_1,n_2;N_1,N_2)=1$.

Next, similar to Eq.~(\ref{eq:R_m}) 
we consider the moments of ${\cal R}(n_1,n_2;N_1,N_2)$.
In this case, because ${\cal R}(n_1,n_2;N_1,N_2)$ has two variables,
we must consider the mixed moments,
\begin{align}
  R_{m_1 m_2}(N_1,N_2) = \sum_{n_1,n_2} n_1^{m_1} n_2^{m_2}
  {\cal R}(n_1,n_2;N_1,N_2) .
  \label{eq:R_mM}
\end{align}
We then Taylor expand $R_{m_1 m_2}(N_1,N_2)$, 
\begin{align}
  R_{m_1 m_2}(N_1,N_2) = \sum_{j_1,j_2} r_{m_1m_2;j_1j_2} N_1^{j_1} N_2^{j_2}.
  \label{eq:TaylorM}
\end{align}
Using Eq.~(\ref{eq:TaylorM}),
the mixed moments of observed particle numbers 
\begin{align}
  \llangle n_1^{m_1} n_2^{m_2} \rrangle
  = \sum_{n_1,n_2} n_1^{m_1} n_2^{m_2} \tilde{P}(n_1, n_2)
  \label{eq:<n1n2>}
\end{align}
are given by 
\begin{align}
  \llangle n_1^{m_1} n_2^{m_2} \rrangle
  = r_{m_1m_2;00} + \sum_{j_1,j_2} r_{m_1m_2;j_1j_2}
  \langle N_1^{j_1} N_2^{j_2} \rangle.
  \label{eq:<>TaylorM}
\end{align}
We note that Eqs.~(\ref{eq:TaylorM}) and (\ref{eq:<n1n2>})
corresponds to Eqs.~(\ref{eq:Taylor}) and (\ref{eq:<>Taylor}), respectively.
In the matrix form, Eq.~(\ref{eq:<>TaylorM}) is expressed as 
\begin{align}
  \left[\begin{matrix}
      \llangle n_1 \rrangle \\ \llangle n_2 \rrangle \\
      \llangle n_1^2 \rrangle \\ \llangle n_1 n_2 \rrangle \\
      \llangle n_2^2 \rrangle \\ \vdots 
  \end{matrix}\right]
  =
  \left[\begin{matrix}
    r_{10;00} \\ r_{01;00} \\ r_{20;00} \\ r_{11;00} \\ r_{02;00} \\ \vdots 
  \end{matrix}\right]
  + \bm{R}_{\rm mult}
  \left[\begin{matrix}
      \langle N_1 \rangle \\ \langle N_2 \rangle \\
      \langle N_1^2 \rangle \\ \langle N_1 N_2 \rangle \\
      \langle N_2^2 \rangle \\ \vdots 
  \end{matrix}\right],
  \label{eq:nNmatM}
\end{align}
where the matrix $\bm{R}_{\rm mult}$ is composed of $r_{m_1m_2;j_1j_2}$.
By inversely solving Eq.~(\ref{eq:nNmatM}) as in Eq.~(\ref{eq:nNmat-inv}),
one obtains the formulas to obtain the (mixed-)moments of the true
distribution function $\langle N_1^{m_1} N_2^{m_2}\rangle$.
The cumulants are then constructed from these moments;
see Refs.~\cite{Nonaka:2017kko,Kitazawa:2017ljq,DeWolf:1995nyp,Broniowski:2017tjq},
for the construction of the cumulants with multi-particle species.
When the expansion Eq.~(\ref{eq:TaylorM}) is not closed,
one may truncate the expansion as in Eq.~(\ref{eq:TaylorL});
the truncation at $j_1+j_2\le L$ allows one to carry out 
the correction up to $L$th order.

This procedure can be generalized to the case with
more than two-particle species in a straightforward manner.

\section{Alternative expansions of $R_m(N)$}
\label{sec:expansion}

In this appendix, we consider modification of the procedure in
Sec.~\ref{sec:DRC}
with the use of alternative expansions of $R_m(N)$.
These procedures give the same final result in principle,
but might be effective in suppressing the accumulation of numerical errors
in practical analyses.

First, instead of Eq.~(\ref{eq:TaylorL}) we consider the Taylor expansion
\begin{align}
  R_m(N) = \sum_{j=0}^L \bar{r}_{mj} (N-N_0)^j
  \label{eq:TaylorN0}
\end{align}
at $N=N_0$ with an arbitrary number $N_0$.
Substituting Eq.~(\ref{eq:TaylorN0}) into Eq.~(\ref{eq:<n^m>}),
one obtains
\begin{align}
  \left[\begin{matrix}
      \llangle n \rrangle \\ \llangle n^2 \rrangle \\ \vdots \\ \llangle n^L \rrangle 
  \end{matrix}\right]
  =
  \left[\begin{matrix}
    \bar{r}_{10} \\ \bar{r}_{20} \\ \vdots \\ \bar{r}_{L0} 
  \end{matrix}\right]
  +
  \bar{\bm{R}}
  \left[\begin{matrix}
      \langle N-N_0 \rangle \\ \langle (N-N_0)^2 \rangle \\ \vdots \\ \langle (N-N_0)^L \rangle
  \end{matrix}\right],
  \label{eq:nNmat0}
\end{align}
with $\bar{\bm{R}}=(\bar{r}_{mj})$.
By choosing $N_0=\langle N \rangle$,
Eq.~(\ref{eq:nNmat0}) allows one to obtain the central moments
$\langle (N-\langle N \rangle)^m \rangle$ directly.

We stress that Eq.~(\ref{eq:TaylorN0}) 
represents the same function as Eq.~(\ref{eq:TaylorL})
with an appropriate replacement between $r_{mj}$ and $\bar{r}_{mj}$.
Therefore, if these parameters are determined accurately 
Eqs.~(\ref{eq:nNmatL}) and (\ref{eq:nNmat0}) give the same final result.
However, they can give different results
within numerical precision in practice.
The use of Eq.~(\ref{eq:nNmat0}) would be advantageous in reducing
the numerical error, as the central moments are more closely
related to cumulants than the standard moments.

Second, it is also possible to use expansions 
motivated by the factorial moments.
One may expand $R_m(N)$ as
\begin{align}
  R_m(N) =& f_{m0} + f_{m1} N + f_{m2} N(N-1) + \cdots
  \nonumber \\
  & + f_{mj} N(N-1)\cdots(N-j+1) + \cdots .
  \label{eq:TaylorF}
\end{align}  
Substituting this expansion into Eq.~(\ref{eq:<n^m>}), one obtains
\begin{align}
  \left[\begin{matrix}
      \llangle n \rrangle \\ \llangle n^2 \rrangle \\ \vdots 
  \end{matrix}\right]
  =
  \left[\begin{matrix}
    f_{10} \\ f_{20} \\ \vdots 
  \end{matrix}\right]
  + \bm{F}
  \left[\begin{matrix}
      \langle N \rangle_{\rm f} \\ \langle N^2 \rangle_{\rm f} \\ \vdots 
  \end{matrix}\right],
  \label{eq:nNmatF}
\end{align}
where $\bm{F}=(f_{mj})$ and 
\begin{align}
  \langle N^j \rangle_{\rm f} &= \langle N(N-1)\cdots(N-j+1) \rangle,
\end{align}
are the factorial moments of $P(n)$.
Equation~(\ref{eq:nNmatF}) provides formulas to obtain factorial
moments $\langle N^m \rangle_{\rm f}$ directly.

One can also expand the factorial moments of ${\cal R}(n;N)$ as 
\begin{align}
  R_{{\rm f}, m}(N) =& \sum_n n(n-1)\cdots(n-m+1){\cal R}(n;N)
  \nonumber \\
  =& \bar{f}_{m0} + \bar{f}_{m1} N + \bar{f}_{m2} N(N-1) + \cdots
  \nonumber \\
  & + \bar{f}_{m2} N(N-1)\cdots(N-j+1) + \cdots,
  \label{eq:TaylorF2}
\end{align}
which enables us to connect $\langle N^m \rangle_{\rm f}$ with
the factorial moments of $\tilde{P}(N)$, $\llangle n^m \rrangle_{\rm f}$,
directly as
\begin{align}
  \left[\begin{matrix}
      \llangle n \rrangle_{\rm f} \\ \llangle n^2 \rrangle_{\rm f} \\ \vdots
  \end{matrix}\right]
  =
  \left[\begin{matrix}
    \bar{f}_{10} \\ \bar{f}_{20} \\ \vdots 
  \end{matrix}\right]
  + \bar{\bm{F}}
  \left[\begin{matrix}
      \langle N \rangle_{\rm f} \\ \langle N^2 \rangle_{\rm f} \\ \vdots 
  \end{matrix}\right], 
  \label{eq:nNmatFF}
\end{align}
with $\bar{\bm{F}}=(\bar{f}_{mj})$.
In the binomial model, $\bar{\bm{F}}$ is given by a diagonal
matrix~\cite{Kitazawa:2017ljq}.
When ${\cal R}(n;N)$ is well approximated by the binomial model,
therefore, the numerical analysis of the inverse matrix of $\bar{\bm{F}}$
would be more stable than that of 
$\bar{\bm{R}}$~\footnote{We thank R.~Holzmann for pointing out 
  this property of the matrix $\bar{F}$.}.

\section{Binomial distribution with fluctuating probability}
\label{sec:binomial-fluc}

In this appendix, we calculate the moments of the
binomial distribution Eq.~(\ref{eq:binomial})
but the probability $p$ is fluctuating, 
i.e. the probability distribution given by
\begin{align}
  P_N(n) = \int_0^1 dp G(p) B(n;p,N) ,
  \label{eq:GB}
\end{align}
where $G(p)$ satisfies $\int_0^1 dp G(p)=1$.

To obtain the moments of $P_N(n)$,
\begin{align}
  \langle n^m \rangle = \sum_n n^m P_N(n),
\end{align}
it is convenient to first calculate 
their factorial moments, which are given by
\begin{align}
  &\langle n^m \rangle_{\rm f}
  =\sum_n n(n-1)\cdots(n-m+1) P_N(n)
  \nonumber \\
  &= \int_0^1 dp G(p) \sum_n n(n-1)\cdots(n-m+1) B(n;p,N)
  \nonumber \\
  &= \int_0^1 dp G(p) N(N-1)\cdots(N-m+1) p^m
  \nonumber \\
  &= N(N-1)\cdots(N-m+1) \langle p^m \rangle_{\rm G},
  \label{eq:n^mf}
\end{align}
where $\langle p^m \rangle_{\rm G} = \int_0^1 dp p^m G(p)$
is the moments of $G(p)$ and in the last equality
we used the relation
\begin{align}
  &\sum_n n(n-1)\cdots(n-m+1) B(n;p,N)
  \nonumber \\
  &= N(N-1)\cdots (N-m+1) p^m.
  \label{eq:FM-B}
\end{align}
The moments of $P_N(n)$ is then obtained by converting
Eq.~(\ref{eq:n^mf})~\cite{Kitazawa:2017ljq}.
Explicit results up to the fourth order are 
\begin{align}
  \langle n \rangle
  =& \langle n \rangle_{\rm f}
  = \langle p \rangle_{\rm G} N ,
  \\
  \langle n^2 \rangle
  =& \langle n \rangle_{\rm f} + \langle n^2 \rangle_{\rm f}
  = \langle p-p^2 \rangle_{\rm G} N + \langle p^2 \rangle_{\rm G} N^2 ,
  \\
  \langle n^3 \rangle =& \langle n \rangle_{\rm f} + 3 \langle n^2 \rangle_{\rm f}
  + \langle n^3 \rangle_{\rm f}
  \nonumber \\
  =& \langle p-3p^2+2p^3 \rangle_{\rm G} N + 3\langle p^2-p^3 \rangle_{\rm G} N^2
  + \langle p^3 \rangle_{\rm G} N^3 ,
  \\
  \langle n^4 \rangle =& \langle n \rangle_{\rm f} + 7 \langle n^2 \rangle_{\rm f}
  + 6 \langle n^3 \rangle_{\rm f} + \langle n^4 \rangle_{\rm f}
  \nonumber \\
  =& \langle p-7p^2+12p^3-6p^4 \rangle_{\rm G} N
  \nonumber \\
  &+ \langle 7p^2-18p^3+11p^4 \rangle_{\rm G} N^2
    \nonumber \\
  &+ 6\langle p^3-p^4 \rangle_{\rm G} N^3
    + \langle p^4 \rangle_{\rm G} N^4.
\end{align}
The same manipulation can be repeated for arbitrary higher orders.
From this derivation, it is clear that the $m$th-order
moments of $P_N(n)$ are given by the $m$th-order polynomial of $N$.

\section{Hypergeometric and beta-binomial distributions}
\label{sec:HG-beta}

In this appendix we summarize the definitions and properties of
the hypergeometric and beta-binomial distributions.

We define the hypergeometric and beta-binomial distributions,
$H(n;N,X,Y)$ and $\beta(n;N,a,b)$, as 
\begin{align}
  &H(n;N,X,Y)
  \nonumber \\
  &= \frac{X!}{n!(X-n)!}
  \frac{(Y-X)!}{(N-n)!(X-n)!}
  \frac{ N! (Y-N)!}{Y!},
  \label{eq:HGD}
\end{align}
and
\begin{align}
  \beta(n;N,a,b) = \int_0^1 dp {\cal B}(p;a,b) B(n;p,N),
  \label{eq:beta}
\end{align}
with the beta distribution
\begin{align}
  {\cal B}(p;a,b) = p^a (1-p)^b / B(a,b),
\end{align}
where $B(a,b)$ is the beta function required for the normalization
$\int_0^1 dp {\cal B}(p;a,b)=1$.

The hypergeometric and beta distributions are given in 
urn models as follows.
First, we consider $N_{\rm w}$ white balls and $N_{\rm b}$ black balls
in an urn, and draw $N$ balls from the urn without returning balls
to the urn in each draw.
Then, the probability distribution of the number of white balls, $n$,
is given by the hypergeometric distribution as 
$H(n;N,N_{\rm w},N_{\rm tot})$ with $N_{\rm tot}=N_{\rm w}+N_{\rm b}$.
Next, we consider successive draws of the balls from the urn.
In each draw, when one draws a white (black) ball, two white
(black) balls are returned to the urn.
After repeating this procedure $N$ times,
the probability distribution to draw $n$ white balls in total
is given by the beta-binomial distribution as
$\beta(n;N,N_{\rm w},N_{\rm b})$.
The distributions of $n$ in both urn models become close to the binomial
distribution $B(n;p,N)$ in the limit $N_{\rm tot}\to\infty$
with fixed $p=N_{\rm w}/N_{\rm tot}$, where $p$ represents the probability
to draw a white ball in a draw.
This means that $H(n;N,X,Y)$ and $\beta(n;N,a,b)$ approach the
binomial distribution in the limit $Y\to\infty$ with fixed $X/Y$ 
and $a\to\infty$ with fixed $a/(a+b)$, respectively.

The cumulants of Eqs.~(\ref{eq:HGD}) and (\ref{eq:beta}),
$\langle n^m \rangle_{\rm c}^{\rm HG}$ and $\langle n^m \rangle_{\rm c}^\beta$,
respectively, up to the fourth order are given by
\begin{widetext}
\begin{eqnarray}
  \langle n \rangle_{\rm c}^{\rm HG}
  &=& Np,
  \label{eq:hyp_ana_1} \\ 
  \langle n^2 \rangle_{\rm c}^{\rm HG}
  &=& \frac{N(N-Y)(-1+p)p}{-1+Y},
  \label{eq:hyp_ana_2} \\
  \langle n^3 \rangle_{\rm c}^{\rm HG}
  &=& \frac{N(2N^{2}-3NY+Y^{2})(-1+p)p(-1+2p)}{(-2+Y)(-1+Y)},
  \label{eq:hyp_ana_3} \\ 
  \langle n^4 \rangle_{\rm c}^{\rm HG}
  &=& [N(N-Y)(-1+p)p(6N^{2}(-1- 6(-1+p)p+Y(1+5(-1+p)p)) \nonumber \\ 
    && -6NY(-1-6(-1+p)p + Y(1+5(-1+p)p)) \nonumber  \\ 
    && + (-1+Y)Y(1+Y(1+6(-1+p)p)))]/\left[(-3+Y)(-2+Y)(-1+Y)^{2}\right],
  \label{eq:hyp_ana_4} 
\end{eqnarray}
\begin{eqnarray}
  \langle n \rangle_{\rm c}^\beta
  &=& Np,
  \label{eq:beta_ana_1}\\
  \langle n \rangle_{\rm c}^\beta
  &=& -\frac{N(N+a+b)(-1+p)p}{1+a+b},
  \label{eq:beta_ana_2}\\
  \langle n \rangle_{\rm c}^\beta
  &=& \frac{N(N+a+b) (2N+a+b)(-1+p)p(-1+2p)}{(1+a+b) (2+a+b)},
  \label{eq:beta_ana_3}\\
  \langle n \rangle_{\rm c}^\beta
  &=& -((N(N+a+b)(-1+p)p(6N^{2}(1+6(-1+p)p+(a+b)(1+5(-1+p)p)) \nonumber \\
  && +6N(a+b)(1+6(-1+p)p+(a+b)(1+5(-1+p)p))+(a+b)(1+a+b) \nonumber \\
  && (-1+(a+b)(1+6(-1+p)p))))/((1+a+b)^{2}(2+a+b)(3+a+b))).
  \label{eq:beta_ana_4}
\end{eqnarray}
\end{widetext}
As one finds from Eq.~(\ref{eq:hyp_ana_2}) (Eq.~(\ref{eq:beta_ana_2})),
the hypergeometric (beta-binomial) distribution is 
narrower (wider) than the binomial distribution with
$\langle n^2 \rangle_{\rm c}=p(1-p)N$.

As the $m$th-order cumulants
are given by the $m$th-order polynomial
in Eqs.~(\ref{eq:hyp_ana_1})--(\ref{eq:beta_ana_4}),
the $m$th-order moments of Eqs.~(\ref{eq:HGD}) and (\ref{eq:beta})
are also given by $m$th-order polynomial.
The values of $r_{mj}$ are obtained as the coefficients of
the moments.

\bibliography{main}

\end{document}